\documentclass[aps,twocolumn,prb]{revtex4}%
\usepackage{amssymb}
\usepackage{amsfonts}
\usepackage{amsmath}
\usepackage{amsmath}
\usepackage{amssymb}
\usepackage{graphicx}%
\begin{document}
\title{Anisotropy of spin splitting and spin relaxation in lateral quantum dots}
\author{Vladimir I. Fal'ko$^{1,2}$, B.L. Altshuler$^{2,3}$, O. Tsyplyatev$^{1}$}
\affiliation{$^{1}$ Physics Department, Lancaster University, Lancaster LA1 4YB, United Kingdom}
\affiliation{$^{2}$ Physics Department, Princeton University, Princeton, NJ 08540}
\affiliation{$^{3}$ NEC-Labs America, Inc., 4 Independence Way, Princeton, NJ 08540}
\date{\today}

\begin{abstract}
Inelastic spin relaxation and spin splitting $\varepsilon_{\mathrm{s}}$ in
lateral quantum dots are studied in the regime of strong in-plane magnetic
field. Due to both g-factor energy dependence and spin-orbit coupling
$\varepsilon_{\mathrm{s}}$ demonstrates a substantial non-linear magnetic
field dependence similar to that observed by R.Hanson \textit{\ et al} [Phys.
Rev. Lett. \textbf{91}, 196802 (2003)]. It also varies with the in-plane
orientation of magnetic field due to crystalline anisotropy of spin-orbit
coupling. Spin relaxation rate is also anisotropic, the anisotropy increasing
with the field. When the magnetic length is less than the 'thickness' of GaAs
dot, the relaxation can be order of magnitude faster for $\mathbf{B}%
\Vert\lbrack100]$ than for $\mathbf{B\Vert}[110]$.

\end{abstract}

\maketitle

Proposals to use electronic spin in quantum dots for quantum information
processing have fuelled extensive studies of spin-orbit (SO) coupling in
heterostructures as means to manipulate the electron spin \cite{Qubit} and as
a source of spin relaxation. Recent theories \cite{Nazarov,AF,Loss} and
experiments \cite{Fujusawa,Kouwenhoven,MarcusSO} suggest that spin relaxation
in quantum dots is strongly suppressed by electron confinement but may be
sped-up by a magnetic field, in particular, by the field parallel to the plane
of the lateral stucture.

It is customary to assume that an in-plane magnetic field couples only to spin
of the electron. In this approximation one can describe spin relaxation in
terms of effective two-dimensional (2D) SO coupling \cite{Nazarov,AF,Loss}.
This aproach can be justified provided that $\lambda_{B}>\lambda_{z}$, where
$\lambda_{B}=\sqrt{c\hbar/eB}$ is magnetic length and $\lambda_{z}$ is the
extent of the subband wave function across the 2D plane. In the oposite limit
that corresponds to a strong magnetic field, $\lambda_{B}\ll\lambda_{z}$,
subbands in a heterostructure transform into bulk Landau levels
(magneto-subbands) thus changing parameters of the effective 2D motion
\cite{FieldRange}. This effect has been observed in optical and FIR
spectroscopy of low-density GaAs/AlGaAs heterostructures \cite{Kukushkin},
resonant tunnelling in double-barrier devices \cite{Maan}, and in quantum
transport characteristics of lateral dots \cite{Marcus}.

In this Letter, we propose a theory of the spin relaxation of electrons in
lateral dots in a strong in-plane magnetic field. The field effect on the
orbital electron motion transforms states in low-density heterostructures into
magneto-subbands. We take into acount this crossover as well as the
Dresselhaus-type spin-orbit coupling in GaAs \cite{Dresselhaus}. We show that
at high fields both the inelastic spin-flip time $T_{1}$ at low temperatures
$kT\ll\varepsilon_{\mathrm{s}}$ and the electron spin splitting $\varepsilon
_{\mathrm{s}}$ depend on the magnetic field orientation with respect to
crystallographic axes, which can be used to distinguish the SO
coupling-induced effects from those caused by a hyperfine interaction with
nuclei. We also present analytical description of $\varepsilon_{\mathrm{s}}$
and $T_{1}$ dependences on both the magnitude and direction of the magnetic field.

The Hamiltonian of electrons in a dot made of a lateral GaAs/AlGaAs structure
grown in direction $\mathbf{l}_{z}=[001]$ can be written as
\begin{align}
\hat{H}_{\mathrm{3D}}  & =\frac{\hat{p}_{z}^{2}+\hat{p}_{X}^{2}+\hat{p}%
_{Y}^{2}}{2m}+V(\mathbf{r})+\frac{g\mu B}{2}\sigma_{X}+\hat{H}_{\mathrm{so}%
},\nonumber\\
\hat{H}_{\mathrm{so}}  & =\gamma\hbar^{-3}\sum\limits_{kij=x,y,z}%
\epsilon^{ijk}\hat{p}_{i}\hat{p}_{j}\sigma_{j}\hat{p}_{i}.\label{H3d}%
\end{align}
In Eq. (\ref{H3d}) we use two systems of in-plane coordinates. Axes $x$ and
$y$ (used in the SO coupling term $\hat{H}_{\mathrm{so}}$) are determined by
cristallographic directions [100] and [010], respectively. Axis $X$ is
directed along the in-plane field $\mathbf{B}=\mathbf{l}_{X}B$, with
$\mathbf{l}_{X}=(l_{x},l_{y},0)$ and $Y$ along $\mathbf{l}_{Y}=(-l_{y}%
,l_{x},0)$. In Eq. (\ref{H3d}) the kinematic, $p_{\alpha}\equiv-i\hbar
\partial_{\alpha}$ and canonical, $\hat{p}_{\alpha}$ momenta are written in
the coordinate system $X$ and $Y$. We use the Landau gauge $\mathbf{A}%
=-(z-a)B\mathbf{l}_{Y}$, so that $\hat{p}_{z}=p_{z}$, $\hat{p}_{X}=p_{X}\ $and
$\hat{p}_{Y}=p_{Y}\;-\frac{e}{c}B(z-a)$, with $a$ to be specified later.

In the spin part of $\hat{H}_{\mathrm{3D}}$, $\mu$ is Bohr magneton, g-factor
in GaAs \cite{Snelling} is $g\approx-0.44+\mathbf{\hat{p}}^{2}g^{\prime}$,
$\epsilon^{ijk}$ is the antisymmetric tensor, and SO coupling constant
$\gamma$ according to Refs. \cite{Jusserand,WeakLoc,footnoteRashba} is
$\gamma=\left(  26\pm6\right)  \mathrm{eV\mathring{A}}^{3}$. $\hat
{H}_{\mathrm{so}}$ in Eq. (\ref{H3d}) is written in the form which guarantees
that it is Hermitian, despite the non-commutativity of operators $\hat{p}_{x}$
and $\hat{p}_{y}$ with $\hat{p}_{z}$ \cite{notations}.

The dot is formed by a potential profile $V=V_{z}(z)+\frac{1}{2}m\vartheta
^{2}(X^{2}+Y^{2})$, which is stronger in the heterostructure growth direction
$\mathbf{l}_{z}$ than within the $XY$ plane. We consider two particular cases:
triangular well $V_{z}=Fz$ and parabolic well $V_{z}(z)=\frac{1}{2}m\omega
^{2}z^{2}$. The wave functions $|n,p_{X},p_{Y}\rangle$ of electrons in the
n-th magneto-subband and their 2D dispersion, $\varepsilon_{n}(p_{Y})+\frac
{1}{2m}p_{X}^{2}$ is determined by
\begin{equation}
\hat{H}_{z}=\tfrac{1}{2m}\hat{p}_{z}^{2}+V_{z}(z)+\tfrac{1}{2}m\omega_{c}%
^{2}\left[  z-a-\lambda_{B}^{2}p_{Y}\right]  ^{2},\label{Hz}%
\end{equation}
whereas the parameter $a$ is chosen in such a way that the lowest subband
dispersion $\varepsilon_{0}(p_{Y})$ has minimum at $p_{Y}=0$. This defines
$\tilde{z}=z-a$ and $m_{Y}^{-1}=\left.  \partial_{p_{Y}}^{2}\varepsilon
_{0}(p_{Y})\right\vert _{p_{Y}=0}$. For a triangular well, we describe
magneto-subbands using the function of a parabolic cylinder
\cite{footnote-functions} and evaluate $a$ and $m_{Y}\equiv\eta m$
numerically. For a parabolic well, harmonic oscillator functions give $a=0$
and $\eta=1+\omega_{c}^{2}/\omega^{2}$.

Since electron confinement across the plane is much stronger than in lateral
directions, $p_{X},p_{Y}\ll p_{z}$, we substitute $\hat{p}_{Y}=p_{Y}%
\;-\frac{e}{c}B\tilde{z}=p_{Y}\;-m\omega_{c}\tilde{z}$, $\hat{p}_{x}%
=l_{x}p_{X}-l_{y}p_{Y}+l_{y}m\omega_{c}\tilde{z}$ and $\hat{p}_{y}=l_{y}%
p_{X}+l_{x}p_{Y}-l_{x}m\omega_{c}\tilde{z}$ into $\hat{H}_{\mathrm{3D}}$,
expand it in powers of kinematic momenta $p_{X}$ and $p_{Y}$ (up to quadratic
terms) and derive the effective 2D Hamiltonian $\hat{H}_{\mathrm{2D}}%
(p_{X},p_{Y},\vec{\sigma})$. In particular, when analysing SO coupling, we
expand $\hat{H}_{\mathrm{so}}$ up to linear order in $p_{Y}$ and $p_{X}$,
\begin{align}
\hat{H}_{\mathrm{so}}  & \approx\hat{H}_{\mathrm{so}}^{0}+\hat{H}%
_{\mathrm{so}}^{1},\;\;\mathrm{where}\label{AA1}\\
\frac{\hat{H}_{\mathrm{so}}^{0}}{\gamma}  & =l_{x}l_{y}\left[  2\dfrac{\hat
{p}_{z}\tilde{z}\hat{p}_{z}}{\hbar^{2}\lambda_{B}^{2}}-\dfrac{\tilde{z}^{3}%
}{\lambda_{B}^{6}}\right]  \sigma_{X}\nonumber\\
& +\left(  l_{x}^{2}-l_{y}^{2}\right)  \dfrac{\hat{p}_{z}\tilde{z}\hat{p}_{z}%
}{\hbar^{2}\lambda_{B}^{2}}\sigma_{Y}+\left(  l_{x}^{2}-l_{y}^{2}\right)
\left(  \dfrac{\tilde{z}\hat{p}_{z}\tilde{z}}{\hbar\lambda_{B}^{4}}\right)
\sigma_{z},\nonumber\\
\frac{\hat{H}_{\mathrm{so}}^{1}}{\gamma}  & =\left(  l_{x}^{2}-l_{y}%
^{2}\right)  \left[  \left(  \dfrac{\hat{p}_{z}^{2}}{\hbar^{2}}-\dfrac
{\tilde{z}^{2}}{\lambda_{B}^{4}}\right)  \frac{p_{X}\sigma_{X}}{\hbar}%
-\dfrac{\hat{p}_{z}^{2}}{\hbar^{2}}\frac{p_{Y}\sigma_{Y}}{\hbar}\right]
\nonumber\\
& -l_{x}l_{y}\left(  \dfrac{2\hat{p}_{z}^{2}}{\hbar^{2}}\allowbreak
+\allowbreak\dfrac{\tilde{z}^{2}}{\lambda_{B}^{4}}\right)  \frac{p_{X}%
\sigma_{Y}}{\hbar}\nonumber\\
& -l_{x}l_{y}\left(  \dfrac{2\hat{p}_{z}^{2}}{\hbar^{2}}-\dfrac{3\tilde{z}%
^{2}}{\lambda_{B}^{4}}\right)  \frac{p_{Y}\sigma_{X}}{\hbar}\nonumber\\
& -\allowbreak\left(  \left[  l_{x}^{2}\allowbreak-l_{y}^{2}\allowbreak
\right]  p_{Y}+2l_{x}l_{y}p_{X}\right)  \allowbreak\dfrac{\tilde{z}\hat{p}%
_{z}+\hat{p}_{z}\tilde{z}}{\hbar^{2}\lambda_{B}^{2}}\sigma_{z}.\nonumber
\end{align}
In both $\hat{H}_{\mathrm{so}}^{0}$ and $\hat{H}_{\mathrm{so}}^{1}%
\allowbreak\allowbreak$ the last term does not contribute to the effective 2D
Hamiltonian: for magneto-subbands determined by $\hat{H}_{z}$ in Eq.
(\ref{Hz}) $\langle0,p_{X},p_{Y}|\tilde{z}p_{z}\tilde{z}|0,p_{X},p_{Y}%
\rangle=0$ and $\langle0,p_{X},p_{Y}|\tilde{z}p_{z}+p_{z}\tilde{z}%
|0,p_{X},p_{Y}\rangle=0$.

The first term in $\hat{H}_{\mathrm{so}}^{0}$ yields an anisotropic addition
to the 2D electron spin splitting linear in $\gamma$. The second term slightly
turns the spin quantization axis off the magnetic field $\mathbf{B=l}_{X}B$.
It can be neglected as long as we restrict ourselves by the lowest order in
$\gamma$. Thus,
\begin{gather}
\varepsilon_{\mathrm{s}}=g\mu B-l_{x}l_{y}\gamma\lambda_{z}^{-3}%
A_{s},\;\;\;\;\label{Es1}\\
g\approx-0.44+\langle0|p_{z}^{2}|0\rangle g^{\prime},\;\lambda_{z}=\left(
\hbar^{2}/mF\right)  ^{1/3},\nonumber\\
A_{s}=\dfrac{\lambda_{z}^{3}\langle0|\tilde{z}^{3}|0\rangle}{\lambda_{B}^{6}%
}-2\dfrac{\lambda_{z}^{3}\langle0|p_{z}\tilde{z}p_{z}|0\rangle}{\lambda
_{B}^{2}\hbar^{2}}.\nonumber
\end{gather}
The anisotropy of spin splitting is crucially sensitive to the inversion
asymmetry of the confinement potential $V_{z}$, thus it is a peculiarity of
heterostructures. The anisotropy effect in Eq.(\ref{Es1}) is maximal in a
field oriented along crystallographic directions $[110]$ or $[1\bar{1}0]$. The
field dependence of the anisotropic part of spin splitting is characterised by
the parameter $A_{s}$. In a weak magnetic field, $\omega_{c}\hbar
<\varepsilon_{1}-\varepsilon_{0}$, perturbation theory analysis gives
$A_{s}\approx2.46\frac{m\lambda_{z}^{2}}{\hbar}\omega_{c}=3.42\omega_{c}%
\hbar/(\varepsilon_{1}-\varepsilon_{0})$ leading to the anisotropy of linear
g-factor. The field dependence $A_{s}(B)$ at high fields is shown in Fig.1(a).
For GaAs/AlGaAs heterostructure with $\lambda_{z}\sim100\mathrm{\mathring{A}}$
and $\gamma\sim\left(  26\pm6\right)  \mathrm{eV\mathring{A}}^{3}$,
Eq.(\ref{Es1}) predicts that the spin splitting $\varepsilon_{\mathrm{s}}$ is
modulated by about 10\% for different orientations of the magnetic field.
$\varepsilon_{s}$ also includes an isotropic non-linear $B$-dependent part due
to the g-factor dependence on the electron momentum, $\langle0|p_{z}%
^{2}|0\rangle\sim Be\hbar/2c$, thus $g\mu B\approx-0.44\mu B+\frac{e\hbar}%
{2c}g^{\prime}\mu B^{2}$. A non-linear $\varepsilon_{\mathrm{s}}%
(B)$-dependence similar to that described by Eq. (\ref{Es1}) was reported in
Ref. \cite{Kouwenhoven} where measurements have been made in a field applied
along $[1\bar{1}0]$ axis \cite{KouwPrivate}.

\begin{figure}[ptb]
\center{\includegraphics[scale=0.33]{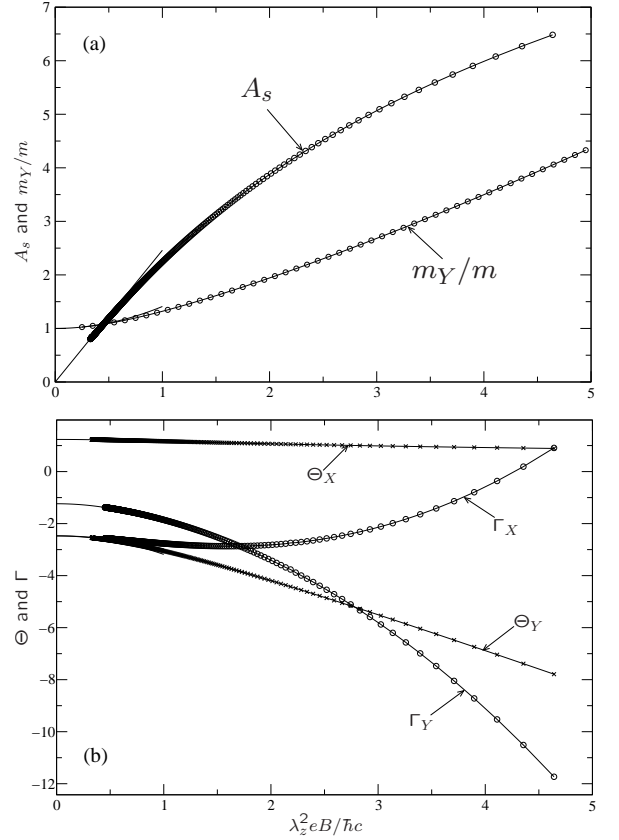}}\caption{Magnetic field
dependence of parameters (a) $A_{s}$, $\eta=m_{Y}/m$ and (b) $\Theta_{X,Y}$,
$\Gamma_{X,Y}$ for potential well $V_{z}=Fz$.}%
\end{figure}


In the effective 2D Hamiltonian,
\begin{align}
\hat{H}_{\mathrm{2D}}  & =\frac{\left(  p_{X}-\hat{a}_{X}\right)  ^{2}}%
{2m}+\frac{\left(  p_{Y}-\hat{a}_{Y}\right)  ^{2}}{2\eta m}\label{H2D}\\
& +\frac{m\vartheta^{2}(X^{2}+Y^{2})}{2}+\frac{\varepsilon_{\mathrm{s}}}%
{2}\sigma_{X},\nonumber\\
\hat{a}_{X}  & =-\hbar\lambda_{\mathrm{so}}^{-1}\left(  \left[  l_{x}%
^{2}-l_{y}^{2}\right]  \Theta_{X}\sigma_{X}+l_{x}l_{y}\Theta_{Y}\sigma
_{Y}\right)  ,\nonumber\\
\hat{a}_{Y}  & =-\hbar\lambda_{\mathrm{so}}^{-1}\left(  l_{x}l_{y}\Gamma
_{X}\sigma_{X}+\left[  l_{x}^{2}-l_{y}^{2}\right]  \Gamma_{Y}\sigma
_{Y}\right)  ,\nonumber
\end{align}
SO coupling arises from the first three terms in $\hat{H}_{\mathrm{so}}^{1}$.
It manifests itself via non-Abelian gauge fields $\hat{a}_{X},\hat{a}_{Y}$
which display an anisotropy linked both to the direction of external magnetic
high field $\mathbf{B=l}_{X}B$ and crystalline axes,%

\begin{align*}
\Theta_{X}  & =\hbar\lambda_{z}^{2}(\kappa-\xi),\;\Theta_{Y}=-\hbar\lambda
_{z}^{2}(2\kappa+\xi),\\
\Gamma_{X}  & =-\hbar\lambda_{z}^{2}(2\kappa-3\xi)\eta,\;\Gamma_{Y}%
=-\hbar\lambda_{z}^{2}\kappa\eta,\\
\lambda_{\mathrm{so}}^{-1}  & =\frac{\gamma m}{\hbar^{2}\lambda_{z}^{2}%
},\;\kappa=\langle0|\frac{p_{z}^{2}}{\hbar^{3}}|0\rangle\;\;\mathrm{and}%
\;\;\xi=\frac{\langle0|\tilde{z}^{2}|0\rangle}{\hbar\lambda_{B}^{4}}.
\end{align*}
Fig. 1(b) shows how $\Theta$ and $\Gamma$ depend on the magnetic field
\cite{footnote-functions} for a triangular well $V_{z}=Fz$ [with $\lambda
_{z}=\left(  \hbar^{2}/mF\right)  ^{1/3}$]. At high fields these dependences
are similar to what we found for a parabolic well $V_{z}(z)=\frac{1}{2}%
m\omega^{2}z^{2}$ with $\lambda_{z}=\sqrt{2\hbar/m\omega}$,%
\[
\Theta_{X}=\frac{1}{\varrho};\Theta_{Y}=\frac{1}{\varrho}-3\varrho;\Gamma
_{X}=\varrho^{3}-3\varrho;\Gamma_{Y}=-\varrho^{3}%
\]
where $\varrho=\sqrt{1+\omega_{c}^{2}/\omega^{2}}\approx\lambda_{z}%
^{2}eB/2\hbar c$ at $\omega_{c}\gg\omega$. This similarity implies that at
high field we can approximate $\Theta,\Gamma$, and $m_{Y}$ in heterostructures
by their values obtained for a parabolic well with the same $\lambda_{z}$.

Lateral orbital states described by $\hat{H}_{\mathrm{2D}}$ have the spectrum
$E_{MM^{\prime}}=(M+\frac{1}{2})\hbar\vartheta+(M^{\prime}+\frac{1}{2}%
)\hbar\vartheta\eta^{-1/2}$. The lowest level wave function
\cite{footnote-functions} is $|\mathbf{0}\rangle=\left(  \pi\lambda\lambda
_{Y}\right)  ^{-1/2}e^{-X^{2}/\lambda^{2}}e^{-Y^{2}/\lambda_{Y}^{2}}%
\varphi_{0}(z)$, where $\lambda=\sqrt{2\hbar/m\vartheta}$ and $\lambda
_{Y}=\eta^{-1/4}\lambda$. Here, $\vartheta$ and $\eta^{-1/2}\vartheta$ are the
frequencies of electron harmonic oscillations along the $X$ and $Y$ axes,
respectively, and the dot states $|\mathbf{n}\rangle=|M,M^{\prime}$ $\rangle$
are characterized by quantum numbers $M$ and $M^{\prime}$.

The rate of the phonon-assisted spin flip $|\mathbf{0},+\rangle\rightarrow
|\mathbf{0},-\rangle$ in the lowest order in both the e-ph interaction and SO
coupling is
\begin{gather}
T_{1}^{-1}=\frac{2\pi}{\hbar}\int\frac{L^{3}d\mathbf{q}}{(2\pi)^{3}}%
|A|^{2}\delta(\varepsilon_{\mathrm{s}}-\hbar sq),\label{AAA}\\
A=\sum_{\mathbf{n}\neq\mathbf{0}}\left[  \frac{\langle\mathbf{0}%
|W|\mathbf{n}\rangle\langle\mathbf{n}|h_{\mathrm{so}}^{Y}|\mathbf{0}\rangle
}{E_{\mathbf{0}}-E_{\mathbf{n}}+\varepsilon_{\mathrm{s}}}+\frac{\langle
\mathbf{0}|h_{\mathrm{so}}^{Y}|\mathbf{n}\rangle\langle\mathbf{n}%
|W|\mathbf{0}\rangle}{E_{\mathbf{0}}-E_{\mathbf{n}}-\varepsilon_{\mathrm{s}}%
}\right]  .\nonumber
\end{gather}
Here, $W(\mathbf{r},\mathbf{q})=w\cdot e^{i\mathbf{qr}}/L^{3/2}$ is the phonon
field with $L^{3}$ being the normalisation volume for phonons. We choose
\[
\left\vert w\right\vert ^{2}=\left(  \frac{\beta^{2}}{q_{\mathrm{s}}}+\Xi
^{2}q_{\mathrm{s}}\right)  ,\;\mathrm{where}\;q_{\mathrm{s}}=\frac
{\varepsilon_{\mathrm{s}}}{\hbar s}.
\]
to take into account both piezoelectric ($\beta$) and deformation ($\Xi$)
phonon potential.

Operators $h_{\mathrm{so}}^{\alpha}(\mathbf{\hat{p}},\mathbf{r})$ can be
obtained using Eq. (\ref{H2D}),%

\begin{equation}
\vec{\sigma}\cdot\vec{h}_{\mathrm{so}}(\mathbf{\hat{p}},\mathbf{r}%
)=-\frac{p_{X}\hat{a}_{X}}{m}-\frac{p_{Y}\hat{a}_{Y}}{m_{Y}}%
.\label{Hso-generic}%
\end{equation}
As long as the orbital part of the Hamiltonian $\hat{H}_{\mathrm{2D}}$ remains
T-invariant, the orbital eigenstates are real, and $\langle\mathbf{0}%
|e^{i\mathbf{qr}}|\mathbf{n}\rangle=\langle\mathbf{n}|e^{i\mathbf{qr}%
}|\mathbf{0}\rangle$. Moreover, $\langle\mathbf{n}|\vec{h}_{\mathrm{so}%
}|\mathbf{0}\rangle=-\langle\mathbf{0}|\vec{h}_{\mathrm{so}}|\mathbf{n}%
\rangle$ because spin operator $\vec{\sigma}$ changes sign under the
$t\rightarrow-t$ transformation whereas the product $\vec{\sigma}\cdot\vec
{h}_{\mathrm{so}}$ remains the same (as a spin-orbit part of T-invariant
$\hat{H}_{\mathrm{2D}}$). Consequently, two terms in the amplitude of the
phonon-emission-assisted spin-flip process in Eq. (\ref{AAA}) cancel
\cite{Nazarov} in the limit $\varepsilon_{\mathrm{s}}\rightarrow0$, and the
transition amplitude reads
\begin{equation}
A=2w\epsilon_{\mathrm{s}}\sum_{M,M^{\prime}\geq1}\frac{\langle\mathbf{0}%
|h_{\mathrm{so}}^{Y}\,|M,M^{\prime}\rangle\langle M,M^{\prime}|e^{i\mathbf{qr}%
}|\mathbf{0}\rangle}{\left(  \hbar\vartheta M+M^{\prime}\hbar\vartheta
\sqrt{m/m_{Y}}\right)  ^{2}-\varepsilon_{\mathrm{s}}^{2}}.\label{Ao}%
\end{equation}
Being generic for any T-invariant Hamiltonian \cite{FootnoteIntersub} such a
cancellation should take place in all orders in $p_{X}$ and $p_{Y}$, hence it
is sufficient to analyse $A$ using only the linear in momentum SO coupling in
$\hat{H}_{\mathrm{2D}}$.

In a parabolic dot operators $p_{X}$ and $p_{Y}$ couple the state
$|\mathbf{0}\rangle=|0,0\rangle$ only to states $|0,1\rangle$, $|1,0\rangle$,
and%
\begin{gather}
\langle0,0|e^{i\mathbf{qr}}|1,0\rangle=\langle1,0|e^{i\mathbf{qr}}%
|0,0\rangle=\frac{i}{2}q_{X}\lambda\Lambda,\nonumber\\
\langle0,0|e^{i\mathbf{qr}}|0,1\rangle=\langle0,1|e^{i\mathbf{qr}}%
|0,0\rangle=\frac{i}{2}q_{Y}\lambda_{Y}\Lambda,\nonumber\\
\Lambda=\langle\mathbf{0}|e^{i\mathbf{qr}}|\mathbf{0}\rangle\approx
f(q_{z})e^{-\frac{1}{8}\left[  \left(  q_{X}\lambda\right)  ^{2}+\left(
q_{Y}\lambda_{Y}\right)  ^{2}\right]  },\label{Lambda}%
\end{gather}
where $f(q_{z})=\int dze^{iq_{z}\tilde{z}}|\varphi_{0}(\tilde{z})|^{2}$. As a
result,%
\begin{equation}
A=\frac{w\epsilon_{\mathrm{s}}}{\hbar\vartheta}\frac{\Lambda\lambda^{2}%
}{2\lambda_{\mathrm{so}}}\left\{  \frac{l_{x}l_{y}\Theta_{Y}q_{X}}{1-\left(
\varepsilon_{\mathrm{s}}/\hbar\vartheta\right)  ^{2}}+\frac{\left[  l_{x}%
^{2}-l_{y}^{2}\right]  \Gamma_{Y}q_{Y}}{1-\left(  \varepsilon_{\mathrm{s}%
}/\hbar\vartheta\right)  ^{2}\eta}\right\}  .\label{AAAA}%
\end{equation}

The angular distribution of the phonon emission is determined by the
form-factor $\Lambda(\mathbf{q})$ in Eq.(\ref{Lambda}). Depending on the
magnetic field, the emitted phonon wavelength $q_{\mathrm{s}}=\varepsilon
_{\mathrm{s}}/\hbar s$ may fall into one of the following regimes:%

\[
\mathrm{\mathrm{A)}}\;q_{\mathrm{s}}<\lambda^{-1};\;\;\;\;\mathrm{B)\;}%
\lambda^{-1}<q_{\mathrm{s}}<\lambda_{Y}^{-1};\;\;\;\;\mathrm{C)\;}\lambda
_{Y}^{-1}<q_{\mathrm{s}}.
\]
In regime A, the phonon wavelength excedes all dimensions of quantum dot. As a
result, $\Lambda\approx1$ and phonons are emitted isotropically. In regime B,
most of phonons are emitted perpendicularly to the magnetic field direction,
since the phonon wavelength is shorter than the lateral dot size $\lambda$ in
the direction of external field. Accordingly, $e^{iq_{z}\tilde{z}}\approx1$,
$e^{-\left(  q_{Y}\lambda/2\right)  ^{2}}\approx1$, thus $f\approx1$ and
$|\Lambda|^{2}\,\approx e^{-\left(  q_{X}\lambda/2\right)  ^{2}}$. Finally, in
the high-field regime C phonons are emitted across the heterostructure. Using
the similarity between magneto-subband states \cite{footnote-functions} and
bulk Landau levels, we approximate $f\approx\exp\left[  -\left(  q_{z}%
\lambda_{B}/2\right)  ^{2}\right]  $ and $|\Lambda|^{2}\,\approx\exp\left[
-\left(  q_{X}\lambda/2\right)  ^{2}-\left(  q_{Y}\lambda_{Y}/2\right)
^{2}-\frac{1}{2}\left(  q_{z}\lambda_{B}\right)  ^{2}\right]  $.

The rate $T_{1}^{-1}$ can be evaluated \cite{footnoteInt} using Eqs.
(\ref{AAA})-(\ref{AAAA}),
\begin{align}
T_{1}^{-1}  & =\frac{\lambda_{\mathrm{so}}^{-2}\,w_{s}^{2}}{4\pi\hbar^{2}%
s}\left(  \dfrac{\varepsilon_{\mathrm{s}}\Gamma_{Y}}{\hbar\vartheta}\right)
^{2}Q\times\label{T1-result}\\
& \times\left\{  \frac{\left(  l_{x}^{2}-l_{y}^{2}\right)  ^{2}}{\left[
1-\eta\varepsilon_{\mathrm{s}}^{2}/\hbar^{2}\vartheta^{2}\right]  ^{2}}%
+\frac{\left(  \alpha l_{x}l_{y}\right)  ^{2}}{\left[  1-\varepsilon
_{\mathrm{s}}^{2}/\hbar^{2}\vartheta^{2}\right]  ^{2}}\right\}  ,\nonumber
\end{align}
where $w_{\mathrm{s}}^{2}=(\beta^{2}/q_{\mathrm{s}})+\Xi^{2}q_{\mathrm{s}}$,
while $Q$ and $\alpha$ are specific for each particular regime (A-C),
\[%
\begin{array}
[c]{ll}%
Q_{\mathrm{A}}=\frac{1}{12}\left(  \lambda q_{\mathrm{s}}\right)  ^{4}, &
\alpha_{\mathrm{A}}=\Theta_{Y}/\Gamma_{Y};\\
Q_{\mathrm{B}}=\frac{\sqrt{\pi}}{8}\left(  \lambda q_{\mathrm{s}}\right)
^{3}, & \alpha_{\mathrm{B}}=2\Theta_{Y}/q_{\mathrm{s}}\lambda\Gamma_{Y};\\
Q_{\mathrm{C}}=(\lambda/\lambda_{Y})^{3}e^{-\frac{1}{2}\left(  q_{\mathrm{s}%
}\lambda_{B}\right)  ^{2}},\;\;\; & \alpha_{\mathrm{C}}=\Theta_{Y}\lambda
_{Y}/\Gamma_{Y}\lambda.
\end{array}
\]

The factor in curly brackets in Eq.(\ref{T1-result}) determines the relaxation
rate dependence on the magnetic field orientation. If the field is so weak
that $\lambda_{z}<\lambda_{B}$, then $\frac{\Theta_{Y}}{\Gamma_{Y}}\approx2$,
$\lambda\approx\lambda_{Y}$, $\alpha=2$, and $T_{1}^{-1} $ turns out to be
isotropic. The anisotropy develops when $\lambda_{B}\sim\lambda_{z}$
[\textit{i.e.}, $\eta>1$, $\lambda>\lambda_{Y}$ and $\Theta_{Y}/\Gamma_{Y}<2$]
and increases with the field. According to Eq. (\ref{T1-result}), at high
fields where $\lambda_{B}\ll\lambda_{z}$ but $\varepsilon_{\mathrm{s}}%
<\hbar\vartheta\ $spin relaxes faster in a magnetic field oriented along
$[100]$ or $[010]$ and slower when $\mathbf{B}$ is parallel to $[110]$ or
$[1\bar{1}0]$, which was the field orientation \cite{KouwPrivate} in the
experiment in Ref. \cite{Kouwenhoven}.\ In the field range where $\lambda
_{B}\ll\lambda_{z}$ and $\varepsilon_{\mathrm{s}}<\hbar\vartheta/\eta^{1/2}$,
spin-flip rate for those two orientations has power-law different field
dependences \cite{fotenoteMaximum}, $T_{1}^{-1}\left(  \mathbf{B}\Vert
\lbrack100]\right)  \propto B^{17/2}$ and $T_{1}^{-1}\left(  \mathbf{B\Vert
}[110]\right)  \propto B^{7/2}$.

The anisotropy in $T_{1}$ is strongly enhanced in the vicinity of crossing of
the level $|\mathbf{0},+\rangle$ with $|0,1,-\rangle$ or $|1,0,-\rangle$,
though the divergence of $T_{1}^{-1}$ in Eq. (\ref{T1-result}) at
$\varepsilon_{\mathrm{s}}=\hbar\vartheta/\eta^{1/2}$ and $\varepsilon
_{\mathrm{s}}=\hbar\vartheta\ $is an artifact of the lowest-order pertirbation
theory analysis and it is prevented by level anti-crossing due to SO coupling
\cite{Loss1}. For $\mathbf{B}\Vert\lbrack100]$ or $\mathbf{B\Vert}[010]$, spin
relaxation is resonantly sped-up when $\varepsilon_{\mathrm{s}}=\hbar
\vartheta/\eta^{1/2}$. For $\mathbf{B}\Vert\lbrack110]$ and $\mathbf{B\Vert
}[1\bar{1}0]$ the rate $T_{1}^{-1}$ acquires a maximum at a higher field where
$\varepsilon_{\mathrm{s}}=\hbar\vartheta$. For samples used in Ref.
\cite{Kouwenhoven} $\hbar\vartheta\approx1$meV,\ thus the crossing of
$|\mathbf{0},+\rangle$ and $|1,0,-\rangle$ levels was beyond the experimental
field range. However, for $\lambda_{z}\sim10$nm and $\hbar\vartheta\sim1$meV
the crossing of levels $|\mathbf{0},+\rangle$ and $|0,1,-\rangle$ should
enhance spin relaxation at $B$ around $15\div20$T if $\mathbf{B}\Vert
\lbrack100]$ or $\mathbf{B}\Vert\lbrack010]$. The formula in
Eq.(\ref{T1-result}) is not exact when $\hbar\vartheta/\eta^{1/2}%
<\varepsilon_{\mathrm{s}}<\hbar\vartheta\varepsilon_{\mathrm{s}}$,
nevertheless, in that field range the anisotropic behavior of $T_{1}%
^{-1}\left(  \mathbf{B}\right)  $ persists, since the spin-flip for
$\mathbf{B}\Vert\lbrack100]$ or $\mathbf{B\Vert}[010]$ is enhanced due to the
openning of additional relaxation channel $|\mathbf{0},+\rangle\rightarrow
|0,1,-\rangle$.

To conclude, we studied the effects of the spin-orbital coupling on the spin
splitting $\varepsilon_{\mathrm{s}}$ and inelatic spin relaxation rate
$T_{1}^{-1}$ in lateral quantum dots at low temperatures $kT\ll\varepsilon
_{\mathrm{s}}$. We found that $\varepsilon_{\mathrm{s}}$ demonstrates a
sizeable non-linearity and anisotropy in its field-dependence, Eq.
(\ref{Es1}). The anisotropy in the spin relaxation, Eq. (\ref{T1-result}) is
predicted to be even stronger: if the magnetic field $\mathbf{B}$ is high,
$\lambda_{B}\ll\lambda_{z}$, the relaxation can be order of magnitude faster
for $\mathbf{B}\Vert\lbrack100]$ than for $\mathbf{B\Vert}[110]$. The latter
feature of the spin relaxation due to SO coupling can be used to distinguish
it from the spin relaxation involving hyperfine interaction with nuclei.

We thank L.Kouwenhoven, D.Loss and C.Marcus for useful discussions. This work
was supported by the EPSRC-Lancaster Portfolio Partnership, INTAS 03-51-6453,
ARO/ARDA (DAAD19-02-1-0039) and DARPA under QuIST programme.


\begin{thebibliography}{99}
\bibitem {footnote-functions}$|n,0,p_{Y}\rangle$ are functions of parabolic
cylinder $Z^{-1}U\left(  -E,\sqrt{2}\left[  \tilde{z}\lambda_{B}^{-1}%
-p_{Y}\lambda_{B}\hbar^{-1}+\lambda_{B}^{3}\lambda_{z}^{-3}\right]  \right)
$, where $E=(\epsilon/\hbar\omega_{c})+\frac{1}{2}\left(  \lambda_{B}%
/\lambda_{z}\right)  ^{6}-\left(  \lambda_{B}/\lambda_{z}\right)  ^{3}%
(\frac{a}{\lambda_{B}}+\frac{p_{Y}\lambda_{B}}{\hbar})$, $\lambda_{z}=\left(
\hbar^{2}/mF\right)  ^{1/3}$ , $\left(  \lambda_{B}/\lambda_{z}\right)
^{3}=\lambda_{B}F/\hbar\omega_{c}$, $Z^{-1}$ stands for normalisation, and the
hard wall boundary condition are used at $\tilde{z}=-a$. The zero-field
subband splitting is $\varepsilon_{1}-\varepsilon_{0}=1.39\hbar^{2}%
/m\lambda_{z}^{2}$. When $\lambda_{B}\ll\lambda_{z}$, magneto-subband wave
functions are close to Landau levels in the bulk, $\varphi_{0}\sim\exp
(-\tilde{z}^{2}/2\lambda_{B}^{2}) $. For $B=0$, $\kappa\approx1.24/\hbar
\lambda_{z}^{2}$ and $\xi=0$.
\bibitem {FootnoteIntersub}The contribution from the inter-subband part of the
second term in $\hat{H}_{\mathrm{so}}^{0}$, Eqs. (\ref{AA1}) is negligible,
unless g-factor was engineered to be anomalously small.
\bibitem {footnoteInt}We take into account that $\int|...|^{2}q_{Y}%
q_{X}d\mathbf{q}=0$ and approximate $(\hbar s/2)\int(2\pi)^{-2}dq_{z}%
\delta(\hbar sq-\varepsilon_{\mathrm{s}})|\Lambda|^{2}\approx\left\{
\begin{array}
[c]{cc}%
\left[  1-(q_{X}/q_{\mathrm{s}})^{2}-(q_{Y}/q_{\mathrm{s}})^{2}\right]
^{-1/2} & \mathrm{(\mathrm{A)}}\\
\,e^{-\left(  q_{X}\lambda/2\right)  ^{2}}\left[  1-(q_{Y}/q_{\mathrm{s}}%
)^{2}\right]  ^{-1/2} & \mathrm{(B\mathrm{)}}\\
\,e^{-\left(  q_{X}\lambda/2\right)  ^{2}-\left(  q_{Y}\lambda_{Y}/2\right)
^{2}-\frac{1}{2}\left(  q_{s}\lambda_{B}\right)  ^{2}} & \mathrm{(C\mathrm{)}}%
\end{array}
\right.  $.
\bibitem {fotenoteMaximum}In GaAs $ms^{2}\sim10\mu$eV, $gm/2m_{e}%
\approx1.5\times10^{-2}$ [$m_{e}$ is the electron mass in vacuum], and
$\frac{1}{2}\left(  q_{\mathrm{s}}\lambda_{B}\right)  ^{2}=\frac{1}%
{2}(gm/2m_{e})^{2}(\hbar\omega_{c}/ms^{2})\ll1$ up to the field $B\sim50$T.
The latter conclusion differs from that in Ref. \cite{Loss} predicting
$T_{1}^{-1}(B)$ to reach maximum at $q_{\mathrm{s}}\sim\lambda_{z}^{-1}$ (In
Ref. \cite{Loss} the orbital effect of in-plane field has been neglected).
\end{thebibliography}
\end{document}